# DEVELOPMENT OF RF REFERENCE LINE FOR THE LINEAR COLLIDER

T.Naito, K.Ebihara, H. Hayano, J.Urakawa, KEK, High Energy Accelerator Organization,
1-1 Oho, Tsukuba, Japan

*Abstract*

The distribution system of the rf reference line for the linear collider is required to supply stable x-band(11.424GHz) rf reference signal over 25km length within 1°(0.25ps) stability. In order to realize the distribution system, an optical fiber link using a phase stabilized optical fiber has been tested. The phase stabilized optical fiber has been employed at LEP, KEKB, etc.. This paper describes the proposed hardware system and result of the preliminary test of the feedback system for the phase stabilization.

## 1 INTRODUCTION

The stabilization of the rf reference line is a key issue for the large scale accelerator. In the case of Japan Linear Collider(JLC), the total distribution length will be 25km. The main accelerating frequency is 11.424GHz and the rf frequency of the Damping ring is 714MHz. The number of the sector, which consists of 8 klystrons, is about 200. The stability of the rf reference line has to be kept 1°(0.25ps) for all sectors. In order to meet the specification, an optical fiber link using a phase stabilized optical fiber(PSOF) with the phase feedback system was proposed and tested. The PSOF is already used at huge scale accelerators, such as LEP, Spring8, KEKB and KEK-ATF[1][2][3][4]. The advantage of the optical fiber link using the PSOF is a high phase stability for a temperature change, a low attenuation even at the x-band frequency and an immunity to electro/magnetic interference. Usually, the rf reference line uses a coaxial cable and a down-converted frequency to reduce the attenuation at the transmission line. In that case, the accelerating frequency is made at each section by up-conversion. The PSOF has a good temperature stability compared to a coaxial cable. An additional phase feedback system can reduce the phase change of the PSOF. If the feedback gain is set to 100 and the phase change of the fiber is less than 100° between two points, it will be kept within 1° stability. To get more accuracy to the beam, a beam based feedback[5][6] will be adoptable which is not mentioned in this paper. This paper describes the characteristics of PSOF and the phase feedback system tested at 509MHz frequency.

## 2 SYSTEM LAYOUT

Fig. 1 shows overall layout of the JLC and the proposed rf reference distribution. The 11.424GHz of rf reference is distributed directly using the PSOF with phase feedback. One phase feedback system between two points consists of a pair of Wave-length Division Multi/demultiplexer(WDM), 1.3μm and 1.5μm wavelength optical links, a phase detector and phase shifters. The optical links using distributed feedback laser transmitters and microwave photo diodes are commercially available for both wavelength, for example, Ortel 3541(1.3μm)/3741(1.5μm) transmitter and 4518 receiver(1.3μm/1.5μm)[7]. A power loss of the optical feedback is shown in Fig.2. The layout of the phase feedback is shown in Fig.2.

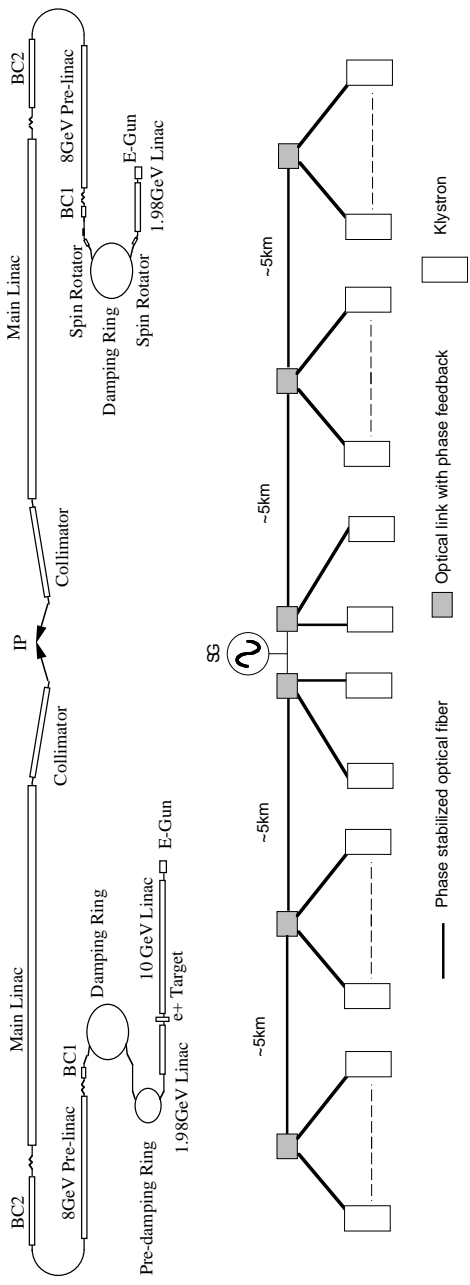

Fig. 1 Overall layout of the JLC and the rf phase reference distribution

fiber is 0.35dB/km for 1.3μm wavelength and 0.20dB/km for 1.5μm wavelength. In order to keep the small power loss and the small phase change, the maximum distance between two points should be kept less than 5km.

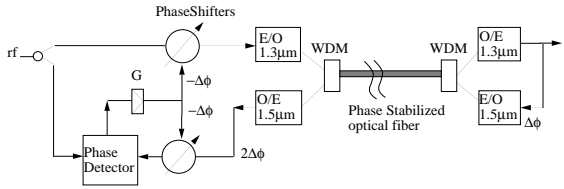

Fig. 2 Phase feedback system using WDM

## 2 DEVICE CHARACTERISTICS

### 2.1 Phase Stabilized Optical Fiber(PSOF)

A normal optical fiber usually has a thermal coefficient of the order of +6ppm/°C. The PSOF is compensated the thermal coefficient by coating a liquid crystal polymer which has negative thermal expansion coefficient[8][9]. The thermal coefficient of the PSOF expansion is less than 0.4ppm/°C. The measured thermal expansion of 500m sample cable of the PSOF is shown in Fig3. The data shows very low thermal expansion at the temperature range from 10°C to 25°C.

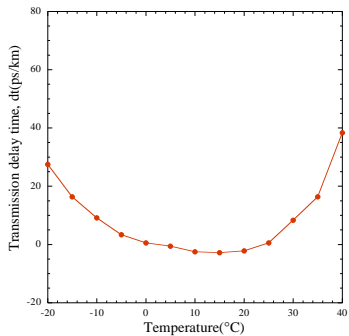

Fig. 3 Temperature characteristics of the PSOF

The thermal expansion of long PSOF cable installed in B Factory Accelerator. The length of the cable is 780m and the 6 fibers are connected in series, then, the total length of the fiber is 4.7km. The cable connects between KEKB injector linac and Center Control Room which goes through the klystron gallery where the temperature is controlled at 25+/-1°C, and some part(~500m) with no temperature control. Fig. 4 shows the phase variation and the temperature in a day. The characteristics according to the temperature of the outside of the building. The phase variation was about 120°/4.7km, which will be reduced when the cable goes through all temperature controlled area like a linear collider tunnel.

### 2.2 Phase feedback

The phase feedback system in optical fiber link was employed in LEP rf reference line[10]. The rf chopping method of the feedback signal was used in the system to avoid mixing of the forward and the backward signals. The system seems to be complicate and difficult to get high stability for the pulsed phase sampling. In order to avoid the signal mixing, we use WDM, which is optical coupler with wavelength selection function. The specification is shown in table 1.

Table 1: Specifications of WDM

|  | **WD-SMF(SWCC)** |
|---|---|
| Operation Wavelength | 1.310μm/1.550μm |
| Band width | 60nm |
| Insertion loss | <1.5dB |
| Isolation | >50dB |
| Directivity | >50dB |

The isolation of the WDM was measured by connecting different frequency for the forward and the backward signal. Fig. 5 shows the power spectrum of the returned signal when connect 509MHz for the forward signal and 508MHz for the backward signal. The isolation ratio was >60dB.

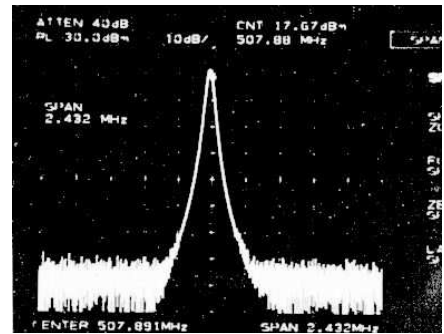

Fig . 5 Isolation measurement of the WDM

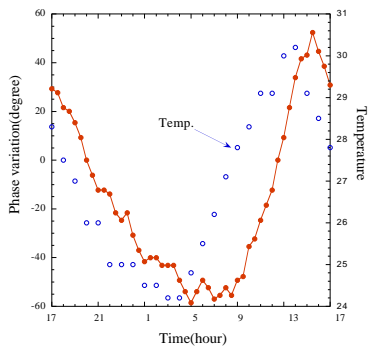

Fig. 4 Characteristics of long cable of the PSOF

The function of the feedback is as follows,
- The input rf reference signal is converted to 1.3μm optical signal, and then the optical signal is transmitted to the other side through PSOF and converted to electric signal. The phase deviation at the point is $\Delta\phi$ which is caused by an expansion of the fiber.
- The signal is returned back through the same PSOF after converted to 1.5μm optical signal. The phase deviation at the returned point is $2\Delta\phi$.

- The detected phase deviation of 2Δϕ is compensated by two phase shifters for each -Δϕ.
- The feedback gain is supplied by the loop gain G. The loop gain is limited by the phase noise mainly comes from the laser diode of the optical link.

### 2.3 Feedback performance test at 509MHz

The feedback using optical link was developed as an upgrade of the KEKB rf phase reference line. The KEKB is a 400m injector linac and two 3km circumference collider rings complex. The acceleration frequency is 509MHz. The present rf reference line of KEKB uses the PSOF without phase feedback and coaxial cables with phase feedback circuit[11]. Each rf station is apart ~1km distance and the PSOF is installed outside of the accelerator to avoid the radiation damage. In order to keep less than 1°(~5ps) stability, the feedback using optical link was developed[12].

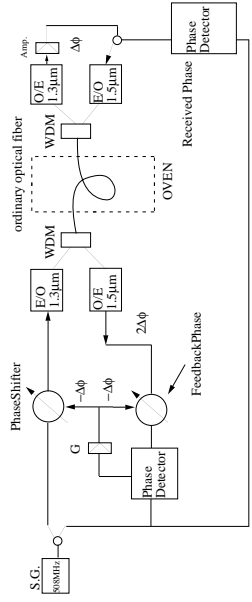

Fig.6 Test circuit of the phase feedback

The test circuit is shown in Fig. 6. The phase change is made by the delay time of an ordinary optical fiber when the temperature is changed. The changed phase is shown as a feedback phase and the compensated phase at opposite side is shown as a received phase. The test result is shown in Fig. 7.

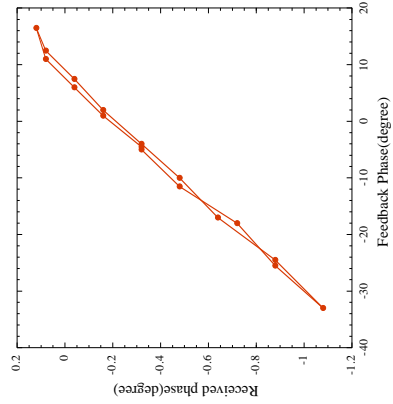

Fig. 7 Performance of the phase feedback

The variation of the received phase should be reducing 1/100 of the feedback phase when the feedback loop gain is set to 100. However the measured feedback gain was ~40. This discrepancy is mainly come from the deviation of the two phase shifters for the control voltage. The phase shifter uses varactor diodes, which has a non-linearity and a strong temperature dependence. The non-linearity is compensated by the linearizer, however the drift of the non-linear curve destroys the compensation. The temperature compensated phase shifter development and total feasibility test is in progress.

## 5 SUMMARY

The development study of the rf reference distribution system for the linear collider was done. The PSOF can be used for long distance signal transmission with less than 120°/5km stability at 25+/-1°C temperature condition. The phase feedback system at 509MHz can reduce the variation ~1/40 or more. The combination of these devices will be useful to make a stable rf reference line. For the next step, the feedback system using x-band frequency is in progress.

## ACKNOWLEDGMENTS

We would like to express our thanks to Professors M.Kihara and K.Takata for their encouragement. We also wish to thank to Dr. Higo and Dr. Mizuno for useful discussion about x-band acceleration system.

## REFERENCES

[1] L. de Jonge et. al., "RF reference distribution for the LEP energy upgrade ", EPAC'94, London, Jul. 1994.
[2] H.Suzuki et. al., "Characteristics of RF Reference Signal Distribution for Spring8", The 9th Symposium on Accelerator Science and Technology, Tsukuba, Japan, Oct. 1993.
[3] T.Naito et. al., "Performance of the Timing System for KEKB", ICALEPCS'99, Trieste, Italy, Oct. 1999.
[4] T.Naito et. al., "Timing System of the ATF", ICALEPCS'97, Beijing, Nov. 1997.
[5] NLC Design group, "Zeroth-Order Report for the Next Linear Collider" SLAC Report 474, May 1996.
[6] A.Gamp et. al., "DESIGN OF THE RF PHASE REFERENCE SYSTEM AND TIMING CONTROL FOR TEH TESLA LINEAR COLLIDER", LINAC'98, Chicago, Aug. 1998.
[7] www.ortel.com
[8] T.Kakuta et. al., "LCP coated fiber with zero thermal coefficient of transmission delay time", International Wire & Cable Symposium, 1987.
[9] J.Urakawa et. al., "The Development of RF Reference Lines and a Timing System for Japan Linear Collider", ICALEPCS'91, Tsukuba, Japan, Nov. 1991.
[10] E. Peschardt et. al., "Phase Compensated Fiber-optic Links for the LEP RF Refrence Distribution", PAC'89, Chicago, Mar. 1989.
[11] K.Ebihara et. al., "RF reference line for KEKB", The 12th Symposium on Accelerator Science and Technology, Wako, Japan, Oct. 1999.
[12] K.Akai et. al., "RF Issues for High Intensity Factories", EPAC'96, Barcerona, June. 1996.